\documentclass[aps,prd,twocolumn,superscriptaddress,amsmath,widetext,amssymbepsfig,showpacs]{revtex4-1}
\usepackage{graphicx}
\usepackage{dcolumn}
\usepackage{bm}
\usepackage{natbib}

\newcommand{\be}{\begin{equation}}
\newcommand{\ee}{\end{equation}}
\newcommand{\bea}{\begin{eqnarray}}
\newcommand{\eea}{\end{eqnarray}}

\newcommand{\neff}{N_{\textrm{eff}}}

\begin{document}
\title{Dark radiation candidates after Planck data}

\author {Eleonora Di Valentino}

\author {Alessandro Melchiorri}
\affiliation{Physics Department and INFN, Universit\`a di Roma ``La Sapienza'', Ple Aldo Moro 2, 00185, Rome, Italy}

\author {Olga Mena}
\affiliation{IFIC, Universidad de Valencia-CSIC, 46071, Valencia, Spain}

\begin {abstract}
Recent Cosmic Microwave Background (CMB) results from the Planck
satellite, combined with previous CMB data and Hubble constant
measurements from the Hubble Space Telescope, provide a constraint on the
effective number of relativistic degrees of freedom 
$3.62^{+0.50}_{-0.48}$ at $95\%$~CL. These new measurements of $\neff$
provide a unique opportunity to place limits
on models containing relativistic species at the decoupling
epoch. Here we review the bounds or the allowed parameter regions in
 sterile neutrino models, hadronic axion models as well as on 
extended dark sectors with additional light species based on the latest Planck CMB observations.
\end {abstract}


\maketitle

\section {Introduction} \label {sec:intro}

Recent Cosmic Microwave Background (CMB) measurements from the Planck
satellite, combined with other cosmological data sets, have provided
new constraints on the  effective number of relativistic degrees of
freedom $\neff$, defined in terms of  the energy density of the total radiation component as
\begin{equation}
 \rho_{rad} = \left[1 + \frac{7}{8} \left(\frac{4}{11}\right)^{4/3}\neff\right]\rho_{\gamma} \, ,
\end{equation}
where $\rho_{\gamma}$ is the current energy density of the CMB. In the standard scenario, the expected value is $\neff=3.046$, corresponding to the three active neutrino contribution and considering effects related to non-instantaneous neutrino decoupling and QED finite temperature corrections to the plasma.
Planck data~\cite{Ade:2013lta}, combined with measurements of the Hubble constant $H_0$
from the Hubble Space Telescope (HST)~\cite{Riess:2011yx} give the constraint
$\neff=3.83\pm 0.54$ at
$95\%$~CL. When low multipole polarization measurements from the Wilkinson Microwave
Anisotropy Probe (WMAP) 9 year data release~\cite{Hinshaw:2012fq}  and high multipole CMB data from both the Atacama Cosmology Telescope
(ACT)~\cite{Sievers:2013wk} and the South Pole Telescope (SPT)~\cite{Hou:2012xq,Story:2012wx} are added in the analysis,
the constraint on $\neff$ is $3.62^{+0.50}_{-0.48}$ at $95\%$~CL~\cite{Ade:2013lta}. These bounds indicate the
presence of an extra dark radiation component at the $\sim 2.4 \sigma$
confidence level. 
Different cosmological analyses carried out previously to Planck data release
including SPT data-only have shown a similar 
evidence~\cite{Archidiacono:2013lva,Calabrese:2013jyk,DiValentino:2013mt}, see also
  Refs. \cite{Hinshaw:2012fq,Giusarma:2012ph,Mangano:2006ur,Hamann:2007pi,Reid:2009nq,Komatsu:2010fb,Izotov:2010ca,Hamann:2010pw,Hou:2011ec,Keisler:2011aw,Smith:2011es,Dunkley:2010ge,darkr,Hamann:2011hu,Nollett:2011aa,Smith:2011ab,darkr2,concha,Diamanti:2012tg,Feeney:2013wp}
  for constraints on the dark radiation abundances exploiting different
  cosmological scenarios, data sets and/or analysis techniques. 
In addition, the presence of an extra dark radiation component will
help enormously in the agreement on the value of the Hubble constant extracted
from CMB Planck data and the value of $H_0$ measured by the HST
team~\cite{Ade:2013lta}. Even if the discrepancy between the CMB and
the astrophysical measurements of $H_0$ can be alleviated in the context of
\textit{Hubble bubble} models~\cite{Marra:2013rba}, in which we, observers, are living inside a local
underdensity, it is mandatory to analyse carefully the constraints
from Planck data on any physical mechanism which
could provide $\Delta \neff \sim 0.6$.

The simplest scenario to explain the extra dark radiation
$\Delta\neff\equiv\neff-3.046$ arising from cosmological data analyses
includes extra sterile neutrino species, since there is no fundamental symmetry in nature
forcing a definite number of right-handed (sterile) neutrino
species. Therefore, sterile neutrinos are allowed in the Standard
Model fermion content. However, there are other possibilities which are as
well closely related to minimal extensions to the standard model of
elementary particles, as thermal axions, or extended dark
sectors with additional relativistic species.  New Planck data provide a unique opportunity to place limits
(or find the favoured regions) on the different parameters which describe the
three models listed above or any other model containing new
light species, see Ref.~\cite{Brust:2013ova}. It is the aim of this paper to carefully
study these limits. Namely, in the case of sterile neutrino
models, the constraints on $\neff$ from recent Planck data can set
upper bounds on the sterile neutrino mixing parameters for
sterile neutrino masses $m \lesssim 0.3$~eV, see Ref.~\cite{Mirizzi:2013kva} for a recent study. We
shall focus here on the so-called (3+1) neutrino mass
models~\cite{3plus1}. In the hadronic axion model~\cite{Kim:1979if,Shifman:1979if}, one can explore, as a function of the axion
mass $m_a$ (being $m_a \lesssim 0.3$~eV) if the axion abundance
(parameterized in terms of $\Delta\neff$) agrees with Planck
findings.  Models containing a dark sector with light species
that eventually decouples from the standard model will also contribute
to $\neff$, as, for instance,  asymmetric dark matter models (see e.g. Refs.~\cite{Blennow:2010qp,Blennow:2012de} and
references therein), or extended weakly-interacting massive
particle models (see the recent work presented in Ref.~ \cite{Franca:2013zxa}). We will follow the expressions from
Ref. ~\cite{Blennow:2012de}, 
in which the authors have followed a general approach to describe the dark sector structure, including 
 both light and heavy relativistic degrees of freedom in the dark
 sector at the time of decoupling. While the former correspond to the number of
degrees of freedom that ultimately constitute the dark radiation sector, the latter
correspond to relatively heavy degrees of freedom that will turn non-relativistic
and heat the dark radiation fluid.  We derive here the constraints on
the number of light and heavy degrees of freedom of the dark sector as
a function of its decoupling temperature from the standard model sector. 

The paper is organised as follows. 
Section~\label {sec:neutr}  presents the constraints
on the sterile neutrino mixing parameters in the ($3+1$) neutrino mass
models. In Section \ref{sec:axion} we
briefly review the thermal axion model and illustrate the constraints on
its mass and its coupling parameter arising  from Planck measurements
on $\neff$. Section \label{sec:adm} analyses the implications from
Planck data on extended dark sectors models. Finally, we draw our conclusions in Sec.~\ref{sec:concl}.

\section {Light sterile neutrino models} \label {sec:neutr}
A number of studies in the literature have been devoted to compute
constraints on the light sterile massive neutrino thermal
abundances~\cite{us,Hamann,Giusarma,Joudaki,latest,Giusarma:2012ph,Archidiacono:2013lva}.
However, the extra sterile neutrinos do not
necessarily need to feature thermal abundances, depending dramatically
their contribution to the mass-energy density 
of the universe on the flavour mixing processes operating
at the decoupling period. Such a study was carried out firstly in
Ref.~\cite{Melchiorri:2008gq}, where the authors computed the
constraints on the
sterile neutrino masses and abundances arising from a joint analysis of short
baseline oscillation and cosmological data.  More recently, the authors of
Ref. ~\cite{Mirizzi:2013kva} have shown that the constraints on $\neff$ from recent Planck data can set
upper bounds on the sterile neutrino mixing angles. 
We benefit here from the approximated expressions provided in
Ref. ~\cite{Melchiorri:2008gq} to explore the constraints on the
sterile neutrino mixing parameters arising from Planck results. The
approximate expressions derived in Ref. ~\cite{Melchiorri:2008gq} are valid here, as we are assuming
small mixing both between the active and heavy
sectors and between the sterile and light neutrino sectors. In other
words, if the flavor
neutrinos $\nu_\alpha$, $\alpha = e, \mu, \tau, s$ (where $s$ refers
to the fourth sterile neutrino)  are related
to the massive base $\nu_i$, $i=1,2,3,4$, through a $4\times4$
unitary matrix which $U$:
\begin{equation}
\nu_\alpha = U_{\alpha i} \nu_i ~,
\label{eq:mixinggen}
\end{equation}
we are assuming that $|U_{a 4}|,|U_{j s}| \ll 1$,
with $a=e, \mu, \tau$ and $j=1,2,3$.
In this case, sterile neutrinos never reach complete thermalization and their
abundances are much lower than the thermal one.
The sterile neutrino contributes to the energy density of the Universe with \cite{Melchiorri:2008gq}:
\be
\Omega_s h^2 \simeq 7 \times 10^{- 5} (\frac{\Delta m_{41}^2}{eV^2}) \sum_a \frac{g_a}{\sqrt C_a} (\frac{U_{a4}}{10^{-2}})^2
\ee
with $a=e, \mu, \tau$ and $\Delta m_{41}^2$ is taken
here as the squared mass of the extra sterile neutrino, 
assuming $m_1\simeq 0$.  The constants $C_a$  ($C_e \sim 0.61$
and $C_{\mu, \tau} \sim 0.17$, respectively) are related
to the effective potential describing the interactions of neutrinos with
the medium.  The constants $g_e\simeq
3.6$ and $g_{\mu}=g_{\tau}\simeq2.5$ are the coefficients of the
damping factor. The contribution from the extra sterile neutrino
to the effective number of relativistic degrees of freedom reads:
\be
\Delta \neff=\frac {\Omega_s h^2}{\frac{7}{8} (\frac{4}{11})^\frac{4}{3} \Omega_{\gamma} h^2}~,
\ee
and therefore, using the Planck measurements of $\Delta \neff$ it is
possible to set constraints on the sterile neutrino mixing parameters,
for a given value of the sterile neutrino mass $m_4$, provided that
$m_4 \lesssim 0.3$~eV. We allow for an electron ($U_{e4}$ ) and muon ($U_{\mu 4}$) flavor content of the sterile
neutrino, setting $U_{\tau 4}=0$. 

\begin{figure*}
\begin{tabular}{c c}
\includegraphics[width=9cm]{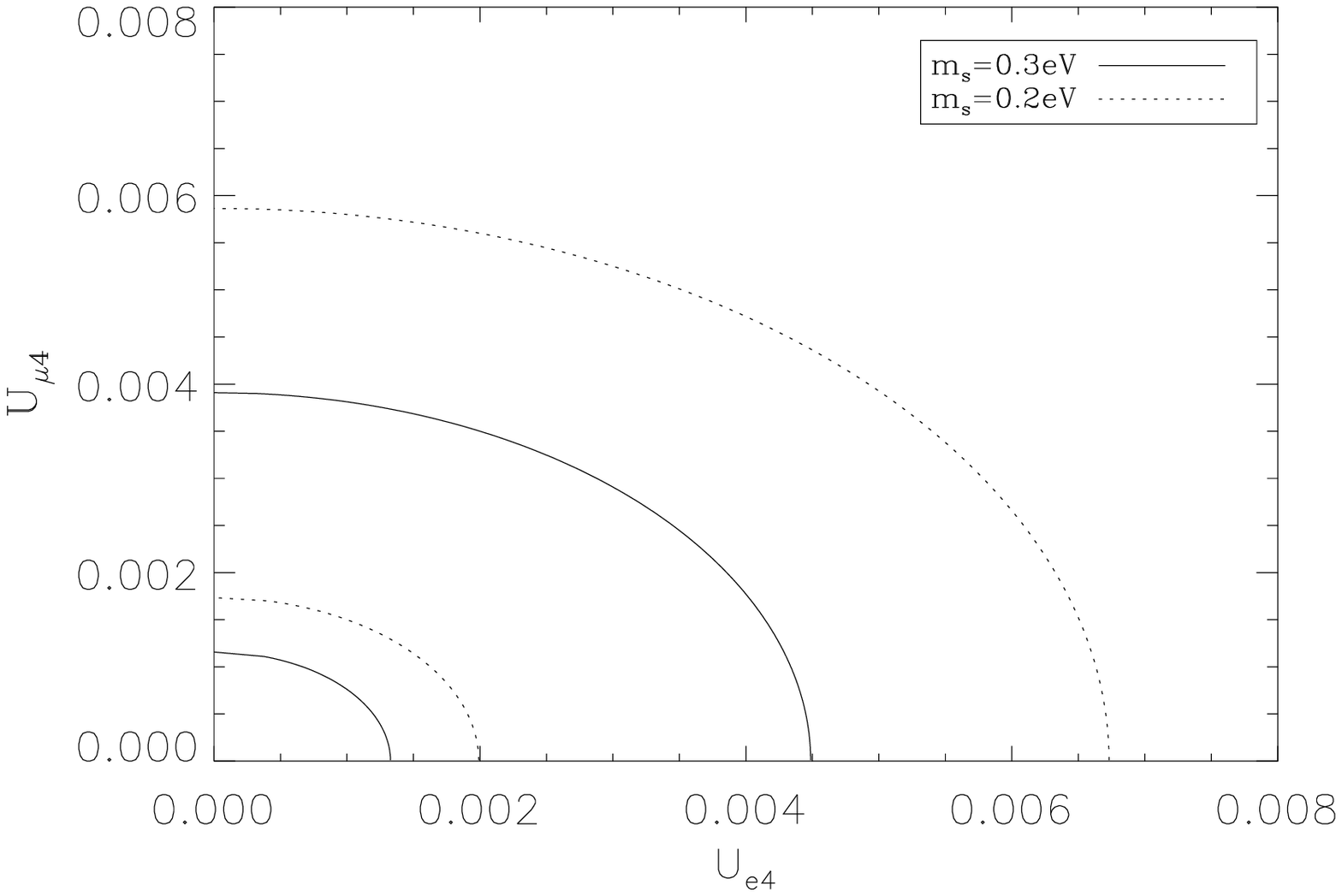}&\includegraphics[width=9cm]{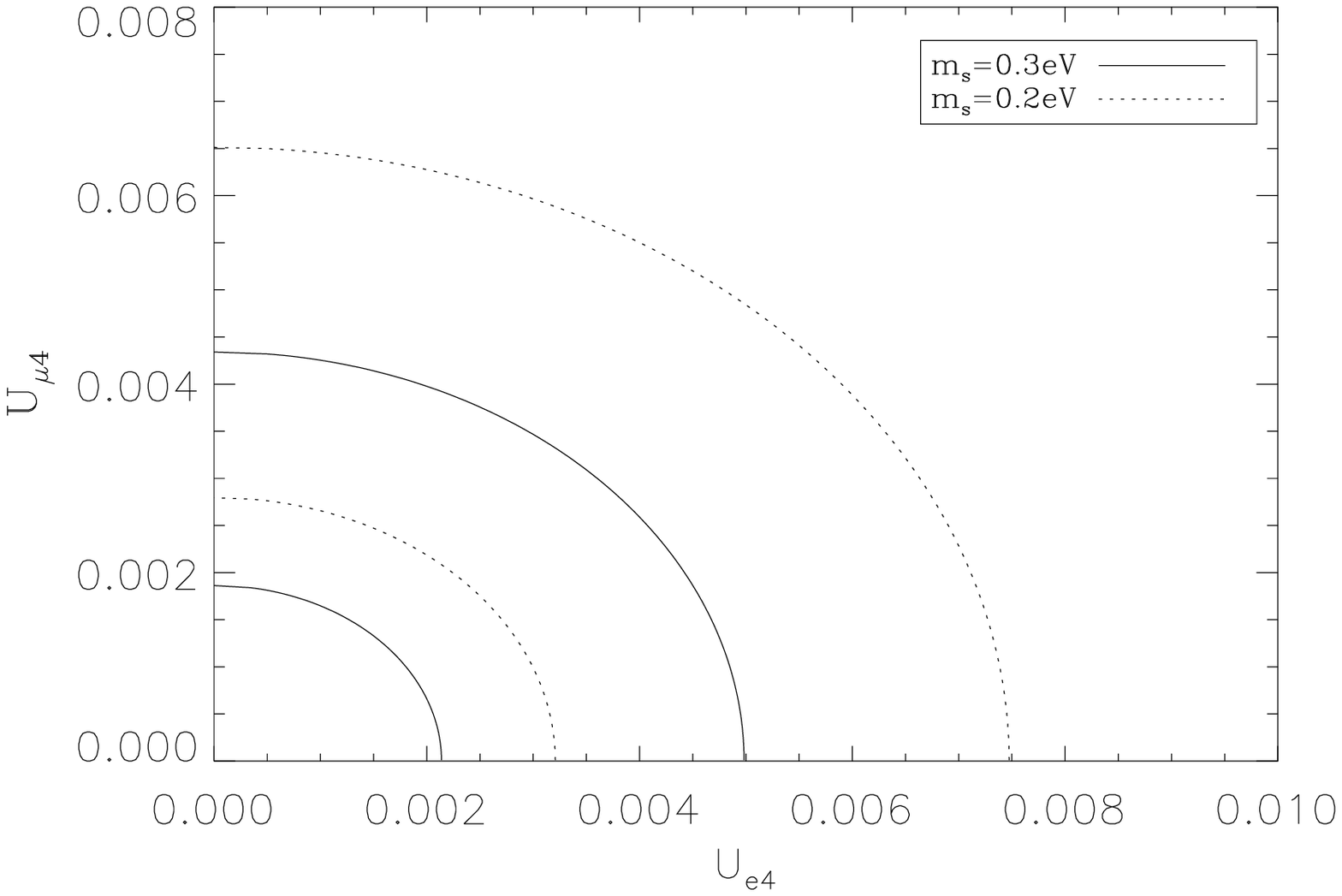}\\
\end{tabular}
 \caption{\small{The left (right) panel illustrates the $95\%$~CL allowed
     regions in the ($|U_{e4}|$,
$|U_{\mu 4}|$) plane arising from the Planck measurements of the
effective number of relativistic degrees of freedom 
   $\neff=3.62^{+0.50}_{-0.48}$ ($\neff=3.83\pm 0.54$) for two values
   of the sterile neutrino mass.}}
\label{sterile}
\end{figure*}

Figure \ref{sterile}, left (right) panel, shows the $95\%$~CL constraints on the ($|U_{e4}|$,
$|U_{\mu 4}|$) plane arising from the Planck constraints, $\neff=3.62^{+0.50}_{-0.48}$
($\neff=3.83\pm 0.54$), for two possible values of the sterile
neutrino mass, $m_s=0.2$ and $0.3$~eV. Larger values of the sterile
neutrino mass will not be relativistic at decoupling and therefore
they can
not be tested exploiting the measured value of $\neff$ by Planck: a
full Montecarlo analysis would be needed, analysis which will be
carried out elsewhere~\cite{inprep}. Notice that the relatively large
values of the sterile neutrino mixing parameters preferred by short
baseline oscillation data in $(3+1)$ models are
excluded here for $0.1 \lesssim m_s \lesssim 0.3$~eV. We find $U_{e4}<0.07$ and $U_{\mu
  4}<0.06$ at the $95\%$~CL for the former range of sterile neutrino
masses. For lower sterile neutrino masses $m_s<0.1$~eV, higher 
mixing parameters are allowed, but such a low sterile neutrino mass is
highly disfavored by oscillation analyses. For instance, the best fit
point to appearance short baseline data in ($3+1$) models is found at 
$\Delta m_{41}^2=0.15$~eV$^2$,  being $U_{e4}=0.39$ and $U_{\mu
  4}=0.39$~\cite{Conrad:2012qt}.  This region of parameters is, however, highly
disfavoured by recent Planck measurements.  Nevertheless one should
keep in mind that the analysis presented here is in the context of
($3+1$) models, which have been shown to be inadequate to fit global data
sets and one should use instead ($3+2$) or ($3+3$) models~\cite{Conrad:2012qt}.

\section{Thermal axion model}\label{sec:axion}
Here we first briefly review the origin of axions. Quantum Chromodynamics (QCD) respects CP symmetry,
despite the existence of a natural, four dimensional, Lorentz and
gauge invariant operator which violates CP. This CP
violating-term will induce a non-vanishing neutron dipole moment,
$d_n$. However, the constraint on the dipole moment $|d_{n}| < 3 \times
10^{-26}\ e$cm~\cite{dipole} requires the CP term contribution to be
 negligible.  Why is CP not broken in QCD?  This is known the so-called
\emph{strong CP problem}.
 The most elegant and promising solution to the strong CP problem was 
provided by Peccei and Quinn~\cite{PecceiQuinn}, by adding a new
global $U(1)_{PQ}$ symmetry, which is spontaneously broken at an
energy scale $f_a$, generating a new spinless particle, the axion. 
The axion mass is inversely proportional to the axion coupling constant $f_{a}$ 
\bea
 m_a = \frac{f_\pi m_\pi}{  f_a  } \frac{\sqrt{R}}{1 + R}=
0.6\ {\rm eV}\ \frac{10^7\, {\rm GeV}}{f_a}~,
\label{eq:massaxion}
\eea
where $R=0.553 \pm 0.043 $ is the up-to-down quark masses
ratio and $f_\pi = 93$ MeV is the pion decay constant. 
Axions can be produced via thermal or non-thermal processes the early
universe, providing a possible (sub)dominant (hot) dark matter
candidate. Here we focus on \emph{hadronic axion models} such as the KSVZ
model~\cite{Kim:1979if,Shifman:1979if}. 

For axion thermalization purposes, only the axion-pion interaction will be relevant.
To compute the axion decoupling temperature $T_D$ we follow the usual freeze out condition
\bea
\Gamma (T_D) = H (T_D)~.
\label{eq:freezeout}
\eea 
The average rate $\pi + \pi \rightarrow \pi
+a$ is given by~\cite{chang}:
\bea
\Gamma = \frac{3}{1024\pi^5}\frac{1}{f_a^2f_{\pi}^2}C_{a\pi}^2 I~,
\eea
where 
\bea
C_{a\pi} = \frac{1-R}{3(1+R)}~,
 \eea
is the axion-pion coupling constant~\cite{chang}, and 
\bea
I &=&n_a^{-1}T^8\int dx_1dx_2\frac{x_1^2x_2^2}{y_1y_2}
f(y_1)f(y_2) \nonumber \\
&\times&\int^{1}_{-1}
d\omega\frac{(s-m_{\pi}^2)^3(5s-2m_{\pi}^2)}{s^2T^4}~,
\eea
where  $n_a=(\zeta_{3}/\pi^2) T^3$ is the number density for axions in
thermal equilibrium, $f(y)=1/(e^y-1)$ denotes the pion distribution
function, 
$x_i=|\vec{p}_i|/T$,  $y_i=E_i/T$ ($i=1,2$), $s=2(m_{\pi}^2+T^2(y_1y_2-x_1x_2\omega))$, and we assume a common mass for the charged and neutral pions, $m_\pi=138$ MeV. 

\begin{figure*}
\begin{tabular}{c c}
\includegraphics[width=9.5cm]{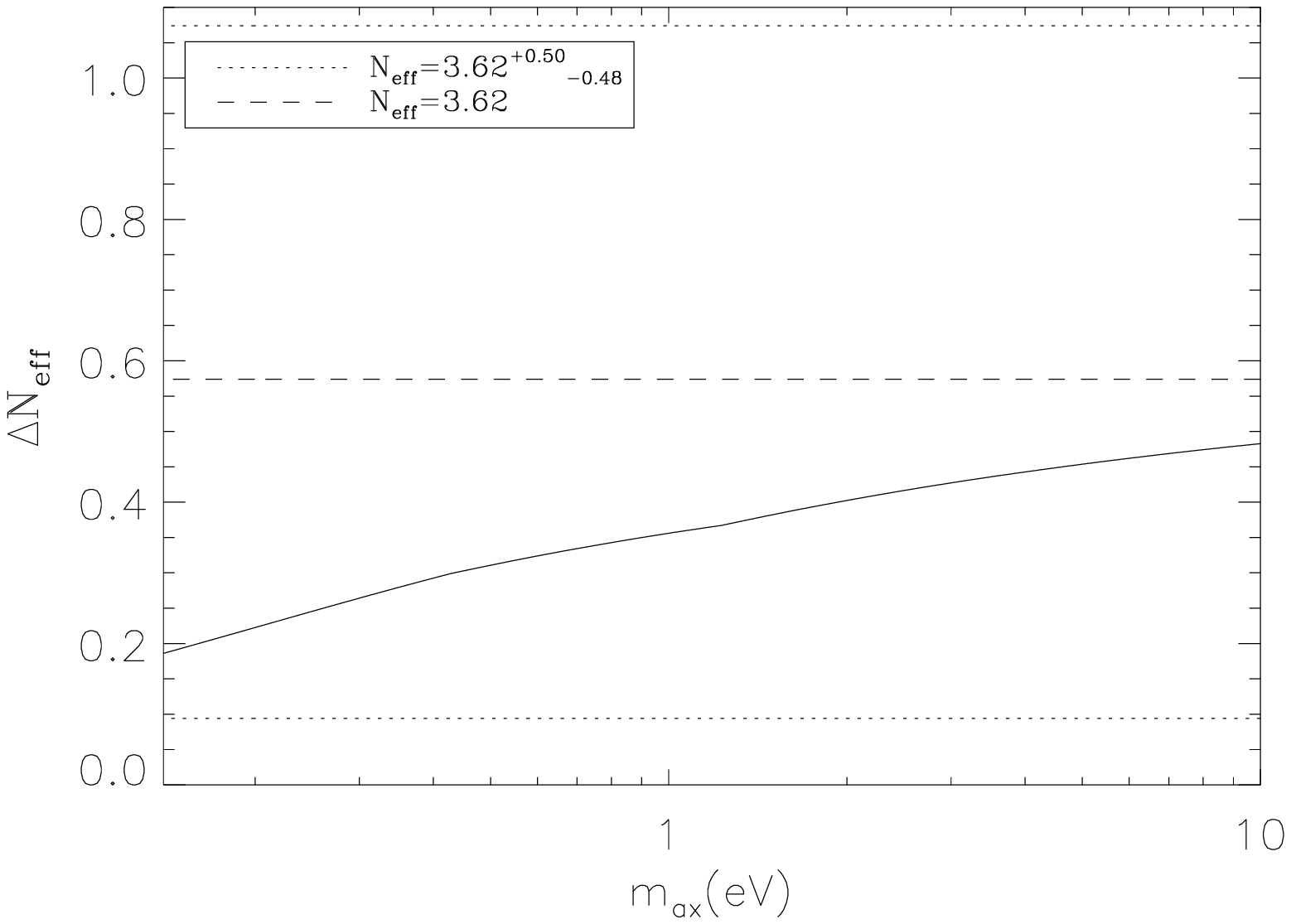}&\includegraphics[width=9.5cm]{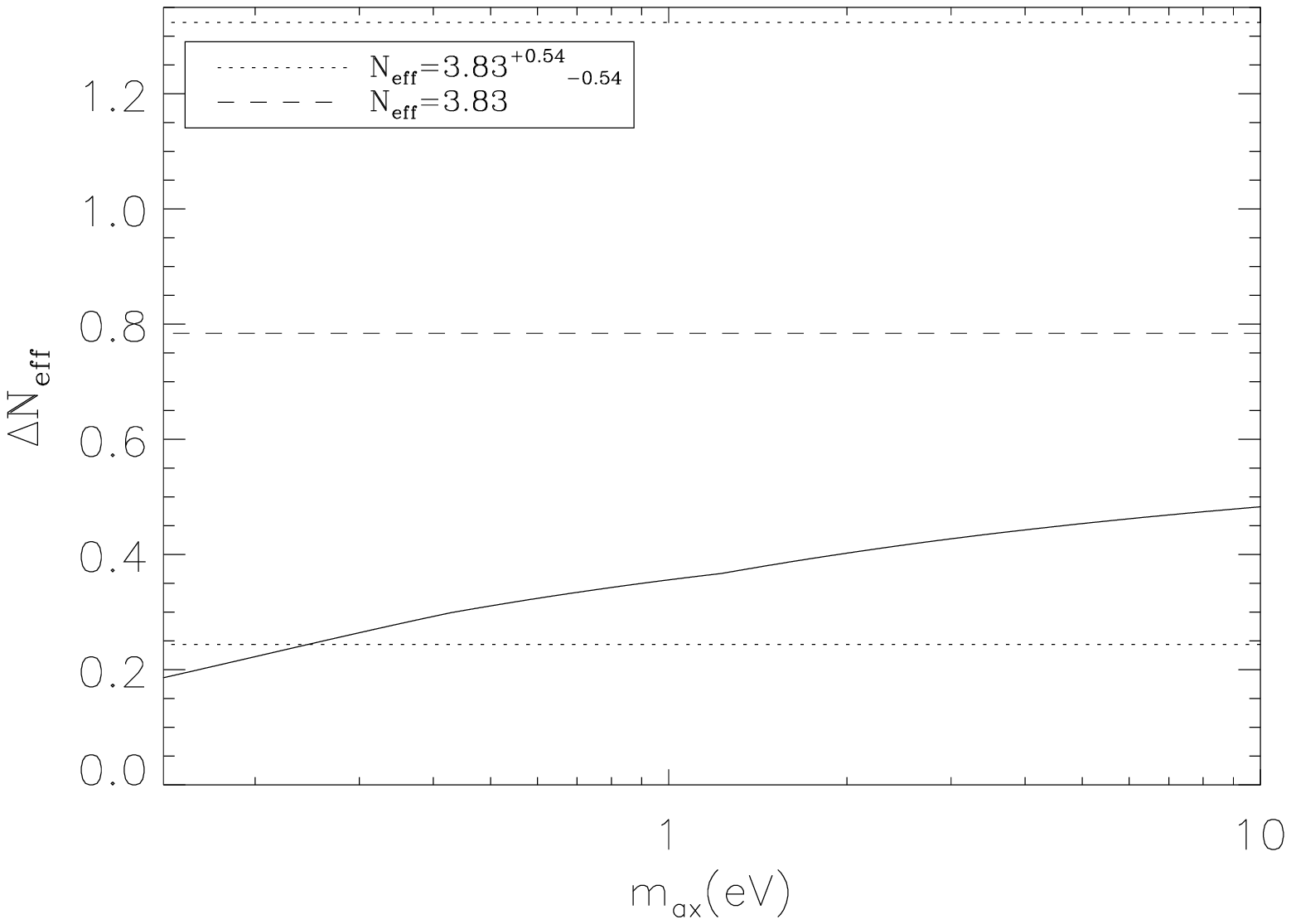}
\end{tabular}
 \caption{\small{$\Delta \neff$ as a function of the thermal axion mass
   (in eV). The left (right) panel illustrates the constraint
   $\neff=3.62^{+0.50}_{-0.48}$ ($\neff=3.83\pm 0.54$).}}
\label{axion}
\end{figure*}

We have numerically solved  the freeze out equation
Eq.~(\ref{eq:freezeout}), obtaining the axion decoupling temperature
$T_D$ versus the axion mass $m_a$ (or, equivalently, versus the axion
decay constant $f_a$). 
From the axion decoupling temperature, we can compute the current axion number density, related to the present photon density $n_\gamma=410.5 \pm 0.5$ cm$^{-3}$ via 
\bea
n_a=\frac{g_{\star S}(T_0)}{g_{\star S}(T_D)} \times \frac{n_\gamma}{2}~, 
\label{eq:numberdens}
\eea  
where $g_{\star S}$ refers to the number of \emph{entropic} degrees of
freedom. At the current temperature, $g_{\star S}(T_0) = 3.91$.
The deviation from the expected value of $\neff$ is $3.046$ due to
the presence of a thermal hadronic axion is given by
\be
\Delta \neff=\frac {\rho_{a}}{\rho_{\nu}}=\frac{4}{3} \left(\frac{3}{2} \frac{n_{a}}{n_{\nu}}\right)^\frac{4}{3}~,
\ee
being $n_\nu$ the current neutrino number density. 
Figure \ref{axion} illustrates the expected $\Delta
\neff$ as a function of the thermal axion mass.  Axions with masses
$m_a \lesssim 0.3$~eV  are still relativistic at the decoupling
epoch and is precisely in this range of values the ones in which CMB $\neff$
measurements can constrain the hadronic axion model.

 If we assume the $\neff=3.83\pm 0.54$, see the right panel of
Fig. \ref{axion}, which corresponds
to the value arising from the combination of Planck data and HST
measurements, the thermal axion model is disfavoured at the
$2\sigma$~CL, since axion masses larger than $0.4$~eV are excluded by 
cosmology~\cite{Hannestad:2005df,Hannestad:2007dd,Melchiorri:2007cd} while axions with masses $m_a < 0.4$~eV do not seem to
provide the appropriate amount of dark radiation, considering Planck
and HST data sets exclusively. However, when other data sets are also added in the analysis, as for instance, WMAP polarization data plus high multipole data from both
ACT and SPT, $\neff$ turns out to be $3.62^{+0.50}_{-0.48}$ and
therefore the hadronic axion model with $m_a\lesssim 0.4$~eV is perfectly compatible with the
value measured of $\neff$ (see the left panel of Fig. \ref{axion}).

\section{Extended Dark sector models}\label{sec:adm}
Any model containing a dark sector with relativistic degrees of
freedom that eventually decouples from the standard model sector will
contribute to $\neff$. An example of these models is the so-called
asymmetric dark matter scenario, which, in general, contains extra radiation degrees of
freedom produced by the annihilations of the thermal dark matter
component. 
We follow here the general approach of Ref. \cite{Blennow:2012de}, in
which the dark sector containts contains 
 both light ($g_\ell$) and heavy ($g_h$)  relativistic degrees of
 freedom at the temperature of decoupling $T_D$  from the standard
 model.  For high decoupling temperature,  $T_D >$ MeV, the
 contribution to the effective number of relativistic degrees of freedom reads~\cite{Blennow:2012de}
\be
\Delta \neff =\frac{13.56}{g_{\star S} (T_D)^\frac{4}{3}} \frac{(g_\ell+ g_h )^\frac{4}{3}}{g_\ell^\frac{1}{3}}~,
\ee
where $g_{\star S} (T_D)$ is calculated using the approximated expression given in Ref. \cite{Wantz:2009it}.
If the dark sector decouples at lower temperatures ($T_D <$ MeV),
there are two possibilities for the couplings of the dark sector
with the standard model: either the the dark sector couples to the
electromagnetic plasma or it couples to neutrinos. In the second case,
which is the one we illustrate here, 
\be
\neff=(3+\frac{4}{7}\frac{(g_h+g_\ell)^\frac{4}{3}}{g_\ell^\frac{1}{3}})(\frac{3\times \frac{7}{4}+g_H+g_h+g\ell}{3 \times \frac{7}{4}+g_h+g_\ell})^\frac{4}{3}~,
\ee
being $g_H$ the number of degrees of freedom  that
become non relativistic between Big Bang Nucleosynthesis and
the dark sector decoupling period.

\begin{figure*}
\begin{tabular}{c c}
\includegraphics[width=9cm]{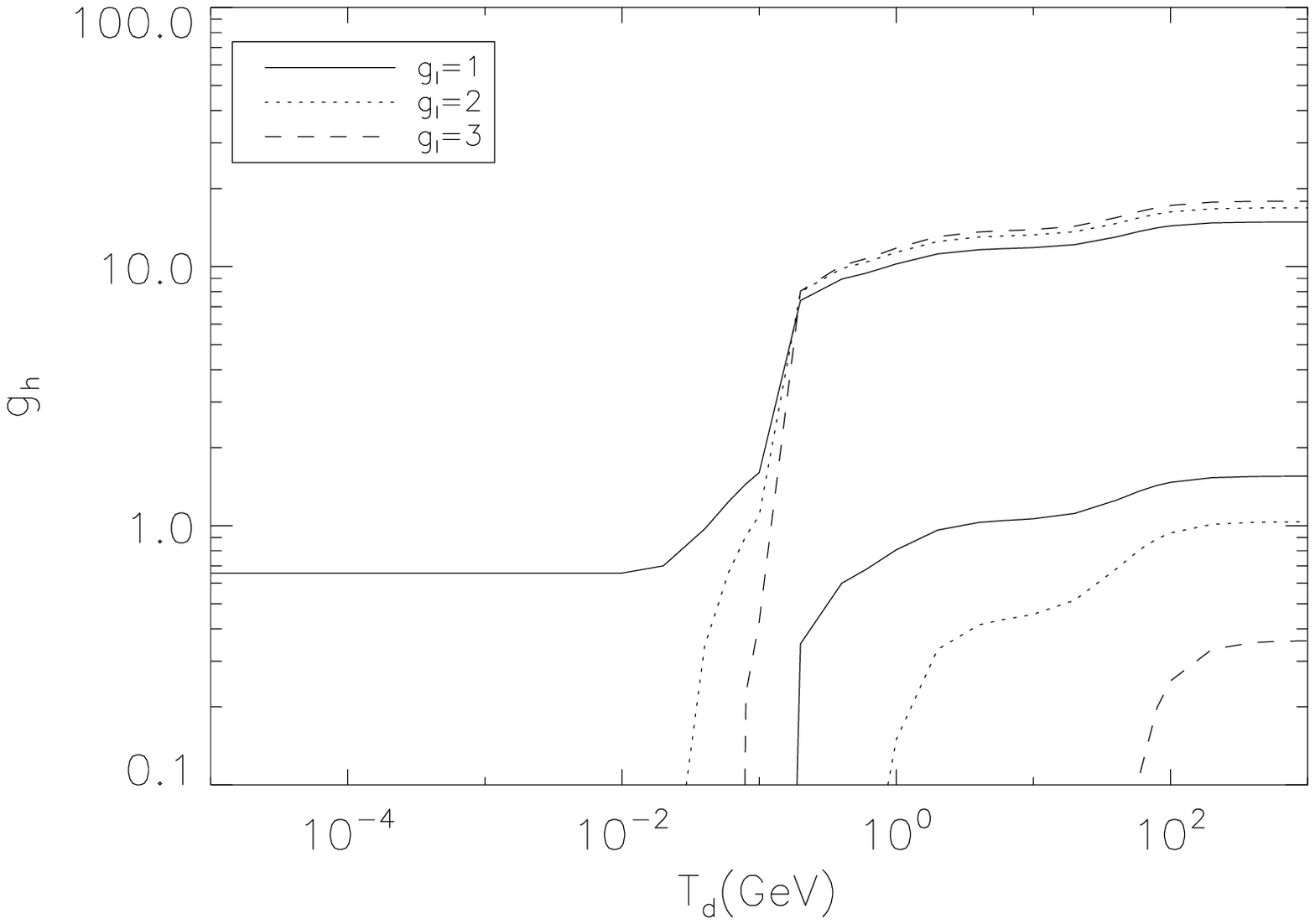}&\includegraphics[width=9cm]{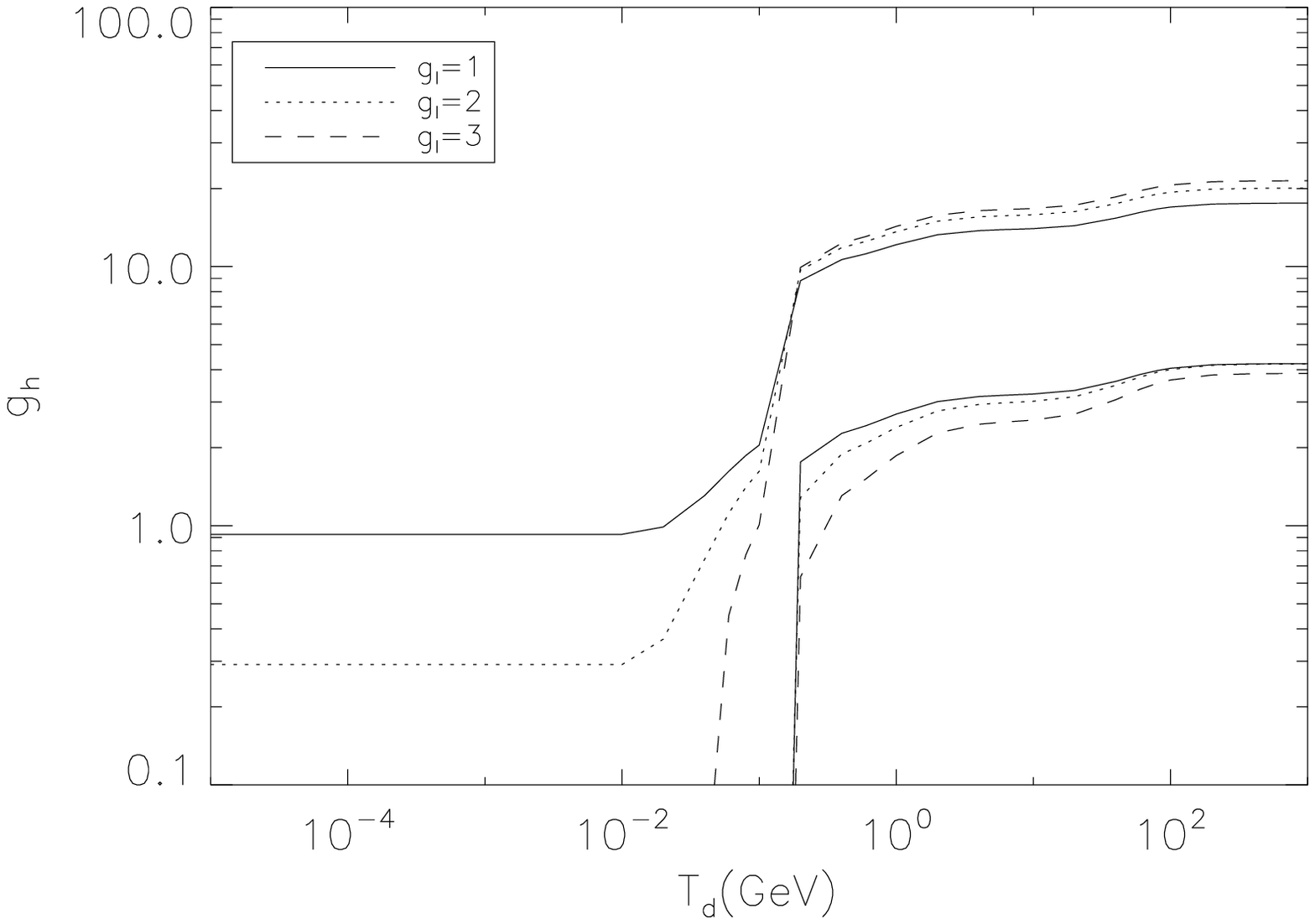}\\
\end{tabular}
 \caption{\small{The left (right) panel shows the $2\sigma$ required ranges for the number of heavy
     degrees of freedom heating the dark sector $g_h$ using $\neff=3.62^{+0.50}_{-0.48}$
($\neff=3.83\pm 0.54$) for several values of $g_\ell$, the light
degrees of freedom of the dark sector.}}
\label{DR_nu}
\end{figure*}

As firstly illustrated in Ref. \cite{Blennow:2012de}, it is possible to
use the measured value of $\neff$ to find the required heavy degrees
of freedom heating the light dark sector
plasma $g_h$ as a function of the dark sector decoupling temperature $T_D$
for a fixed value of $g_\ell$. Figure~\ref{DR_nu}, left (right) panel, illustrates the
$2\sigma$ required ranges for $g_h$ using $\neff=3.62^{+0.50}_{-0.48}$
($\neff=3.83\pm 0.54$), for $g_H=0$.  Notice that at
decoupling temperatures $T_D>$~MeV, the standard model relativistic degrees of
freedom will he heated, requiring therefore heating in the dark sector 
to  enhance the value of $\Delta \neff$. On the other hand, at low
decoupling temperatures, the number of the required heavy degrees of
freedom $g_h$ decreases as $\Delta \neff$ does. Indeed, for the case
of $\neff=3.62^{+0.50}_{-0.48}$ ($\neff=3.83\pm 0.54$), having extra heavy degrees of freedom
is highly (mildly) disfavoured. This is because at low temperatures,
the photon background can not get extra heating from standard model
particles and therefore an extra heating in the dark sector will
increase dramatically the value of $\neff$. 

\section {Conclusions} \label {sec:concl}

Recent Cosmic Microwave Background measurements from the Planck
satellite, combined with measurements of the Hubble constant 
from the Hubble Space Telescope (HST) have provided the constraint
$\neff=3.83\pm 0.54$ at
$95\%$~CL. If low multipole polarization measurements from the Wilkinson Microwave
Anisotropy Probe 9 year data release and high multipole CMB data from both the Atacama Cosmology Telescope
 and the South Pole Telescope are added in the analysis,
the constraint on $\neff$ is $3.62^{+0.50}_{-0.48}$ at $95\%$~CL. These bounds indicate the
presence of an extra dark radiation component at the $\sim 2 \sigma$
confidence level and can be exploited to set limits on any model
containing extra dark radiation species, as sterile neutrino models,
hadronic axion scenarios or extended dark sector schemes.

Within the ($3+1$) sterile neutrino scenario,  we find that the relatively large
values of the sterile neutrino mixing parameters preferred by short
baseline oscillation data in $(3+1)$ models are excluded here for $0.1
\lesssim m_s \lesssim 0.3$~eV. For lower sterile neutrino masses $m_s<0.1$~eV, higher 
mixing parameters are allowed, but such a low sterile neutrino mass is
highly disfavored by oscillation analyses. However, other sterile
neutrino models, as the ($3+2$) or the ($3+3$) scenarios, may provide
 a much better fit to both cosmological measurements
and short baseline data. In the context of the hadronic axion model, the constraint
$\neff=3.83\pm 0.54$ disfavours the former model at the
$2\sigma$~CL. On the other hand, the axion model studied here with
$m_a\lesssim 0.4$~eV is perfectly compatible with cosmological data
when lower values of $\neff$ are considered, as those obtained when
other data sets are analysed together with Planck data. Concerning 
models with a dark sector with light species that eventually decouples
from the standard model, as, for instance, asymmetric dark matter models, having extra heavy degrees of freedom
in the dark sector is highly (mildly) disfavoured for the case of
$\neff=3.62^{+0.50}_{-0.48}$ ($\neff=3.83\pm 0.54$).
Future Planck polarization data will help in cornering dark radiation models.

\subsection*{Acknowledgements}
We would like to thank Enrique Fern\'andez Mart\'{\i}nez for useful comments on the manuscript.
O.M. is supported by the Consolider Ingenio project CSD2007-00060, by PROMETEO/2009/116, by the Spanish Ministry Science project FPA2011-29678 and by the ITN Invisibles PITN-GA-2011-289442.

\label{lastpage}
\end{document}